\documentclass{PoS}

\usepackage{url}
\usepackage{wasysym}

\newcommand{\ket}[1]{\ensuremath{\left|{#1}\right\rangle}}

\newcommand{\desy}{{\sc Desy}}

\title{Dihadron production in semi-inclusive DIS from transversely polarized protons}

\ShortTitle{Dihadron production in SIDIS from transversely polarized protons}

\author{\speaker{S. Gliske}${}^{ab}$ and L. Pappalardo${}^c$ (on behalf of the HERMES Collaboration)\\
\llap{${}^a$} High Energy Physics Division,\\
        Argonne National Laboratory\\
        Lemont, Illinois, USA\\
\llap{${}^b$} University of Michigan\\
        Ann Arbor, Michigan, USA\\
\llap{${}^c$} Dipartimento di Fisica e Scienze della Terra\\
        Universit\`a di Ferrara\\
        44122 Ferrara, Italy\\
E-mail: \email{sgliske@umich.edu}, \email{pappalardo@fe.infn.it}}

\abstract{Transverse Momentum Dependent (TMD) dihadron production,
  including vector meson production, allows access to various TMD
  distribution and fragmentation functions. Dihadron production is
  complementary to single hadron semi-inclusive DIS measurements,
  pairing the same distribution functions with different fragmentation
  functions. While dihadrons present unique measurement opportunities,
  the TMD dihadron cross section is significantly more complex than
  that for single hadron production, due to the polarization in the
  final state. Various theoretical advances, which further clarify the
  complexity, will be highlighted. The HERMES analysis of the
  transverse target moments of the TMD dihadron cross section allows
  the first test of a particular prediction of the Lund/Artru string
  fragmentation model, specifically that the favored Collins
  fragmentation function has opposite sign in single hadron production
  versus vector meson production. The status and results of this
  analysis will be discussed and an extension of the Lund/Artru model
  for disfavored fragmentation will also be presented.  }

\FullConference{XXI International Workshop on Deep-Inelastic Scattering and Related Subject -DIS2013,\\
		22-26 April 2013\\
		Marseilles,France}

\begin{document}

\section{Introduction}

Semi-inclusive deep-inelastic scattering (SIDIS) accesses combinations
of distribution functions and fragmentation functions
\cite{diehlBacchetta}.  SIDIS dihadron production, constituting both
resonant and non-resonant hadron pairs, is not only complimentary to
SIDIS pseudo-scalar production but also provides unique
opportunities.  For example, measurements from both the dihadron and
pseudo-scalar transverse momentum dependent (TMD) cross
section are needed to test the Lund/Artru string fragmentation model
\cite{Artru}.  Additionally, dihadron production allows collinear
access to transversity, which is not possible with single
pseudo-scalar production.  Collinear-based global fits are advantageous
as the evolution equations are not known in the TMD case.

\section{Fragmentation Models}

The Lund/Artru string fragmentation model posits that, as a
transversely polarized struck quark is exiting the proton, a gluon
flux tube breaks into a quark, anti-quark pair with quantum numbers
equal to that of the vacuum, $0^{++}$.  Angular momentum conservation
implies that if the produced anti-quark has spin parallel to the
struck quark (i.e. a $\ket{\frac{1}{2}}\ket{\frac{1}{2}}$ or
$\ket{\frac{1}{2}}\ket{\frac{1}{2}}$ state), the produced hadron will
prefer moving towards the left, while if the spins are anti-parallel
(i.e. a $\ket{\frac{1}{2}}\ket{-\frac{1}{2}}$ or
$\ket{-\frac{1}{2}}\ket{\frac{1}{2}}$ state), the produced hadron will
move towards the right.
Relating the direct sum basis with the direct product basis results in
the conclusion that pseudo-scalar mesons $\left(\ket{0,0} \propto
\ket{\frac{1}{2}}\ket{-\frac{1}{2}} +
\ket{-\frac{1}{2}}\ket{\frac{1}{2}}\right)$ will prefer moving towards
the right while the two transversely polarized vector meson states
$\left(\ket{1,\pm1} = \ket{\pm\frac{1}{2}}\ket{\pm\frac{1}{2}}\right)$
will prefer moving towards the left, implying the Collins function
has opposite sign for these two cases.

Note however, that the Lund/Artru model assumes that the struck quark
is in the produced hadron, typical for favored fragmentation but not
for disfavored fragmentation.  A model for disfavored fragmentation,
denoted the Gliske Gluon Radiation model, can be obtained by assuming
the hadron is produced from the quark, anti-quark pair produced from
the vacuum \cite{myDissertation}, i.e. the struck quark emitting a
high energy, off-shell gluon and the quark returns to the proton
remnant.  Further soft-interactions which do not effect the spin set
additional quantum numbers.
A common sub-diagram is shared between the Lund/Artru model and this
gluon radiation model.  Applying the same angular momentum and spin
states arguments to this model results in the prediction for the
Collins function sign as does the Lund/Artru model.  Although these
models are mainly focused on first rank fragmentation, one can
assume higher rank fragmentation to be mostly negligible in
determining the sign of the Collins function.  Thus, the combination
of these two models suggests that the Collins function for both
favored and disfavored transverse $\left(\ket{1,\pm1}\right)$ vector
meson states have the same sign, opposite to the sign of the Collins
function for favored single pseudo-scalar hadron production
\cite{hermesCollins}.

\section{Theory}

The Lund/Artru and Gliske Gluon Radiation string fragmentation models are both amplitude
level models, not cross section level which involves the amplitude
times its complex conjugate.  To relate the models with the cross
section, a modified definition of the distribution functions is
introduced \cite{myDissertation,mySpin2010}, where the quark,
quark-conjugate spins define the fragmentation function labels and all
final-state polarization is subsumed in the partial wave expansions.
This new convention merits a new partial wave analysis including the
full $\ell$ and $m$ specification of the final-state polarization
\cite{myDissertation,mySpin2010}.

In particular, the collinear interference fragmentation function
$H_1^{\varangle}$ \cite{BRspecModel} is associated with the
$\ket{\ell=1,m=1}$ partial wave of the Collins function, while the
$\ket{2,\pm2}$ partial waves of the Collins function are those which
are expected to have opposite sign as the pseudo-scalar Collins
function.  The new partial wave expansion also highlights the existing
fact that the $H_1^{\varangle}$, i.e. $H_1^{\ket{1,1}}$, receives
contributions from both $sp$ and $pp$ interference.

This new partial wave analysis also capitalizes on the inherent
symmetries in the SIDIS cross section.  Specifically, the cross
section without any partial wave expansion has identical form to the
cross section for producing a pure scalar final state, i.e. single
pseudo-scalar production.  Thus, one can compute the cross section for
any final-state polarization, at any twist, by 1) obtaining the cross
section for single pseudo-scalar production at the desired
twist-level, 2) setting the non-expanded, polarized final-state cross
section to have identical form, and 3) expanding the fragmentation
functions according to spherical harmonics.

Using this method, the dihadron cross section, including twist-2 and
twist-3, for all final-state polarizations has been computed.  The
transverse-target moments were published in Ref. \cite{mySpin2010},
the unpolarized moments were discussed in the presentation
corresponding to these proceedings, and the publication of the
complete expressions is in progress.  The collinear $\ket{1,1}$ moment
related to the Collins function and transversity coincide with the
$\sin(\phi_R+\phi_S)\sin\vartheta$ modulation, while the TMD
$\ket{2,\pm2}$ moments coincide with the
$\sin((1\mp2)\phi_h \pm 2 \phi_R + \phi_S)\sin^2\vartheta$
modulations.  The angle definitions are as in Ref. \cite{mySpin2010}.

The remainder of this document focuses on the analysis of
$\pi^\pm\pi^0$ and $\pi^+\pi^-$ dihadrons, subsuming the
$\rho$-triplet resonances.  Note, the convention is that for a
$h_1~h_2$ di-meson dihadron, assuming $h_1$ and $h_2$ do not have
equal charge, $h_1$ refers to the charged meson when one of the mesons
is neutral, and $h_1$ refers to the positively charged meson when both
$h_1$ and $h_2$ are charged.


\section{Analysis}

Data was taken with the HERMES detector using the HERA 27.6 GeV lepton
beam during the years 2002-2005 with a transversely polarized hydrogen
target.  Lepton-hadron separation efficiency was better than 98\%, and
hadron identification was made using an event-level algorithm and the
ring-imaging Cerenkov (RICH) detector.  Events were required to pass
data quality and SIDIS identification cuts ($Q^2>1$~GeV${}^2$, $W^2 >
10$~GeV${}^2$, $0.023 < x < 0.4$, $0.2 < y < 0.95$, $0.2 < z < 0.8$,
$0.05$~GeV~$<~P_{h\perp}~<~1.60$~GeV).
The $\pi^0$-mesons were identified as a resonant peak in the diphoton mass
spectrum, and the $\pi^\pm\pi^0$ data were corrected for non-resonant
diphoton pairs.  The charge symmetric background and exclusive
background were determined to be negligible.

The effect of the {\sc Hermes} acceptance on the angular variables was
corrected in the parameter space of the amplitudes rather than in the
histogrammed yield space of Ref. \cite{cosnphi}.  Both approaches are
mathematically equivalent, but the parameter-space approach is
desirable in higher dimensions---dihadrons analysis involves four
angular variables and currently uses 42 angular moments in the fit
function (though more moments are present in the cross section), while
pseudo-scalar hadron analysis only involves one or two angular variables (depending on the target polarization).  The 42 moments
used in this analysis are the 24 twist-2 and twist-3 unpolarized
moments and the 18 (twist-2) Collins and Sivers transverse target
moments.

The acceptance correction required the development of a new
Monte Carlo generator, \texttt{TMDGen}, as well as a new TMD spectator
model for dihadron fragmentation.  Both \texttt{TMDGen} and the
spectator model are detailed in Refs. \cite{myDissertation,mySpin2010}.
Smearing and radiative effects are considered in the systematic
uncertainty, but no corrections for those effects were applied.
Systematic uncertainties also consider the variation in results
between the running period with positron instead of electron beams, as
well as uncertainties related to the RICH hadron identification.

\section{Results}

\begin{figure}
  \begin{center}
    \includegraphics[width=0.95\textwidth]{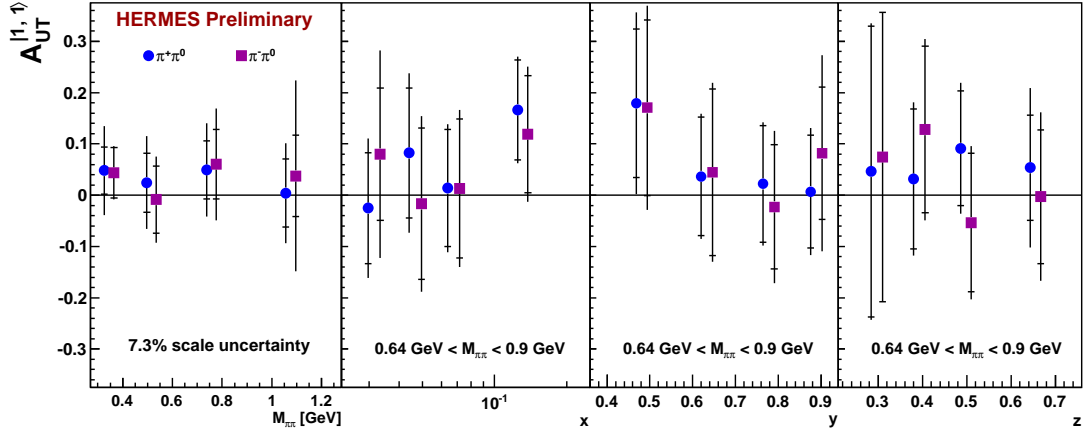}
  \end{center}
  \caption{\label{fig11}The $\sin(\phi_R+\phi_S)\sin\vartheta$
    amplitudes for $\pi^\pm\pi^0$ dihadrons.  The left panel shows the
    results versus the invariant mass of the $\pi\pi$ system,
    $M_{\pi\pi}$, while the right three panels show the $x$, $y$, and
    $z$ dependence of the data restricted to the mass bin centered at
    the $\rho$-mass.  Full error bars are the statistical
    uncertainty (inner error bar) added in quadrature with the
    systematic uncertainty. }
\end{figure}

The results for $\pi^\pm\pi^0$ dihadron $\ket{1,1}$ amplitude related
to transversity and the Collins function are shown in Figure
\ref{fig11}.  The $\pi^+\pi^-$ results are not shown, as they were
previously published \cite{hermesDihadron} and the new analysis uses a
more optimized choice of binning.  The signs of the amplitudes are
consistent across all three $\pi\pi$ dihadrons, although the
statistics are more limited for the $\pi^\pm\pi^0$ dihadrons than for
the $\pi^+\pi^-$ dihadrons.  These results are suitable to be included in
global, collinear extractions of transversity and may further
constrain the flavor separation.

\begin{figure}
  \begin{center}
    \includegraphics[width=0.6\textwidth]{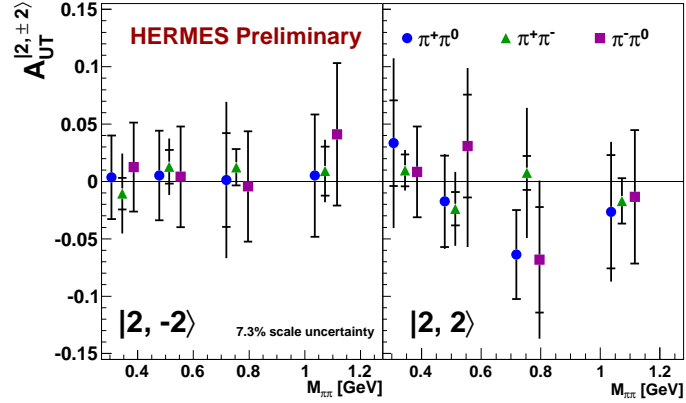}
  \end{center}
  \caption{\label{fig22}The
    $\sin(3\phi_h-2\phi_R+\phi_S)\sin^2\vartheta$ (left panel) and
    $\sin(-\phi_h+2\phi_R+\phi_S)\sin^2\vartheta$ (right panel) amplitudes
    for $\pi\pi$ dihadrons versus the invariant mass of the $\pi\pi$ system.  Error bars are plotted as in Figure 1.}
\end{figure}

The results for $\ket{2,\pm 2}$ Collins amplitudes are shown in Figure
\ref{fig22}.  The results are consistent with zero outside of the
$\rho$-mass peak for both $\ket{2,\pm 2}$ amplitudes, suggesting that the
non-resonant background has no observable presence in this partial
wave.  The $\ket{2,-2}$ amplitude is also consistent with zero in the
$\rho$-mass peak.  One explanation is that the transversity
distribution function causes the struck quark to be in the up-state,
causing the $\ket{2,-2}$ state to be inaccessible and thus zero.  The
$\ket{2,2}$ amplitude in the $\rho$-mass peak for $\pi^\pm\pi^0$
dihadrons are indicate a non-zero, negative value.  The results
are indeed consistent with the expectations of the Lund/Artru and
Gluon Radiation models and encourage repeating these
measurements at existing or future facilities.

\acknowledgments
We gratefully acknowledge the \desy\ management for its support, the staff
at \desy\ and the collaborating institutions for their significant effort,
and our national funding agencies and the EU FP7 (HadronPhysics3, Grant
Agreement number 283286) for financial support.

\end{document}